 \documentclass [12pt,a4paper      ]{article}
\usepackage{times}

\DeclareFontFamily{OT1}{times}{}
\DeclareFontShape {OT1}{times}{m }{n }{ <-> ptmr }{}
\DeclareFontShape {OT1}{times}{bx}{n }{ <-> ptmb }{}
\DeclareFontShape {OT1}{times}{m }{it}{ <-> ptmri}{}
\DeclareFontShape {OT1}{times}{bx}{it}{ <-> ptmbi}{}

\usepackage{amsmath}
\usepackage{amsfonts}
\usepackage{amssymb}
\usepackage{latexsym}

\newcommand{\MU}{ {\mu} }
\newcommand{\NU}{ {\nu} }
\newcommand{\DEF}{:=}   
\newcommand{\rmd}{d} 
\begin{document}

\title{\bf\vspace{-2.5cm} The locally-conserved current density of 
                             the Li\'enard-Wiechert field }

\author{
         {\bf Andre Gsponer}\\
         {\it Independent Scientific Research Institute}\\ 
         {\it Oxford, OX4 4YS, England}
       }

\date{ISRI-06-03.11 ~~ December 31, 2008}

\maketitle

\begin{abstract}

The complete charge-current density and field strength of an arbitrarily accelerated relativistic point-charge are explicitly calculated.  That current includes, apart from the well-established delta-function term which is sufficient for its global conservation, additional contributions depending on the second and third proper-time derivatives of the position.  These extra contributions are necessary for the local conservation of that current, whose divergence must vanish \emph{identically} even if it is a distribution, as is the case here.  Similarly, the field strength includes an additional delta-like contribution which is necessary for obtaining this result.  Altogether, the Li\'enard-Wiechert field and charge-current density must therefore be interpreted as nonlinear generalized functions, i.e., not just as distributions, even though only linear operations are needed to verify local charge-current conservation.

~\\

\noindent 03.50.De ~ Classical electromagnetism, Maxwell equations

\end{abstract}

\section{Introduction}
\label{int:0} \setcounter{equation}{0}

Local conservation of the charge-current density four-vector $J_\MU$ is a necessary consequence of the antisymmetry of the electromagnetic field strength tensor $F_{\mu\nu}$.  Indeed, the identity
\begin{align}\label{int:1}
    \partial^\MU J_\MU \equiv 0,
\end{align}
derives from taking the four-divergence of both sides of Maxwell's inhomogeneous equation
\begin{align}\label{int:2}
    \partial^\nu F_{\mu\nu} =  - 4\pi J_\MU,
\end{align}
where the left-hand side vanishes after contraction because $F_{\mu\nu}$ is antisymmetric.  The vanishing of the four-divergence of the current density is thus not an ordinary conservation law (i.e., a `weak' conservation law subject to the field equations being satisfied) but an \emph{identity} (i.e., a so-called strong conservation law) which has to be satisfied even if $J_\MU$ is a distribution rather than a smooth function.  This conclusion is absolutely general and should therefore be true for an arbitrarily moving relativistic point-charge, that is for the charge-current density of the Li\'enard-Wiechert field, which turns out not to be the case for the customary formulation of this field.

  In this letter we show that if the Li\'enard-Wiechert charge-current density is properly calculated, which implies that the Li\'enard-Wiechert field-strength must be supplemented by an additional $\delta$-function-like field, local charge-current conservation is restored.  This conclusion is obtained by using only well known physical concepts, and a few basic results of distribution theory, but at the expense of some lengthy calculations whose details are given in another paper \cite{GSPON2006D}, to which we refer for all results presented here.

\section{Definitions}
\label{def:0} \setcounter{equation}{0}

Let $Z_\MU$ be the four-position of an arbitrarily moving relativistic point-charge, and $X_\MU$ a point of observation.  At the point $X_\MU$ the four-potential $A_\MU$, field $F_{\mu\nu}$, and four-current density $J_\MU$ are functions of the null interval $R_\MU$ between $X_\MU$ and $Z_\MU$, i.e.,
\begin{align}\label{def:1}
    R_\MU \DEF X_\MU - Z_\MU,
   \qquad\mbox{such that}\qquad
    R_\MU R^\MU = 0,
\end{align}
as well as of the four-velocity $\dot{Z_\MU}$, four-acceleration  $\ddot{Z_\MU}$, and four-biacceleration $\dddot{Z_\MU}$ of the charge, to which three invariants are associated: $\dot{Z}_\MU R^\MU, \ddot{Z}_\MU R^\MU$, and $\dddot{Z}_\MU R^\MU$.  The first one is called the retarded distance,
\begin{align}\label{def:2}
          \xi \DEF   \dot{Z}_\MU R^\MU,
\end{align}
which enables to introduce a `unit' null four-vector $K_\MU$ defined as
\begin{align}\label{def:3}
          K_\MU(\theta,\phi) \DEF R_\MU/\xi,
\end{align}
and the so-called acceleration and biacceleration invariants defined as
\begin{align}\label{def:4}
      \kappa \DEF  \ddot{Z}_\MU K^\MU,
  \qquad\mbox{and}\qquad
        \chi \DEF \dddot{Z}_\MU K^\MU.
\end{align}

    The derivations of $F_{\mu\nu}$ from the four-potential, i.e.,
\begin{align}\label{def:5}
   F_{\mu\nu} = \partial_\MU A_\NU - \partial_\NU A_\MU,
\end{align}
and of $J_\MU$ according to \eqref{int:2}, require that the partial derivatives are calculated at the the position $X_\MU$ under the condition $R_\MU R^\MU = 0$, which insures causality.  For an expression $E = E(X_\MU,\tau)$, where the argument $X_\MU$ corresponds to an explicit dependence on $X_\MU$, and $\tau$ to the proper time, this condition leads to the covariant differentiation rule
\begin{align}\label{def:6}
   \partial_\MU E(X_\MU,\tau) = \partial_\MU E(X_\MU)
                              + K_\MU \dot{E}(\tau) \Bigr|_{\tau=\tau_r}. 
\end{align}
In this equation the condition  $\tau=\tau_r$ implies that all quantities are evaluated at the retarded proper time $\tau_r$.  In the following, for simplicity,  this condition will be specified explicitly only for the main equations.

\section{The customary formulation}
\label{cus:0} \setcounter{equation}{0}

The potential of the Li\'enard-Wiechert field field can be obtained by a number of methods that are explained in many text books, e.g., \cite{SOMME1951-,LANDA1975-,JACKS1975-,BARUT1964-,INGAR1985-}. Most frequently it is obtained by working in the Lorentz gauge, and by means of a Green's function assuming that the point-charge can be represented by a three-dimensional $\delta$-function.  In the covariant notation of this letter this source charge-current density can be written in the following form   
\begin{align}\label{cus:1}
   J^{\rm S}_\MU = \frac{e}{4\pi} \dot{Z}_\MU\frac{1}{\xi^2} \delta(\xi),
\end{align}
whose normalization corresponds to the global (or integral) form of charge conservation, i.e.,
\begin{align}\label{cus:2}
  \int_{0}^{2\pi}\rmd\phi \int_{0}^{\pi}
                 \rmd\theta \sin\theta \int_{0}^{\infty}
                 \rmd\xi~ \xi^2 J_\MU =   e \dot{Z}_\MU.
\end{align}
The resulting potential has the remarkable simple form
\begin{align}\label{cus:3}
   A^{\rm LW}_\MU = e \frac{\dot{Z}_\MU}{\xi} \Bigr|_{\tau=\tau_r},
\end{align}
from which, applying the rule \eqref{def:6}, it is easily found that
\begin{align}\label{cus:4}
   F^{\rm LW}_{\mu\nu} =  e \Bigl[ 
            \frac{K_\MU\ddot{Z}_\NU}{\xi}
            + (1-\kappa\xi)\frac{K_\MU \dot{Z}_\NU}{\xi^2}
            - \{ \mu \leftrightarrow \nu \}
                   \Bigr]_{\tau=\tau_r},
\end{align}
where $\{ \mu \leftrightarrow \nu \}$ means two more terms with $\mu$ and $\nu$ interchanged.  Using again the rule \eqref{def:6} the corresponding charge-current density calculated according to \eqref{int:2} is then
\begin{align}\label{cus:5}
   J^{\rm LW}_\MU = \frac{e}{4\pi} \Bigl[\frac{\dot{Z}_\MU}{\xi^2}
                     +2 \frac{\ddot{Z}_\MU - \kappa\dot{Z}_\MU}{\xi}
                 \Bigr] \delta(\xi) \Bigr|_{\tau=\tau_r}.
\end{align}
This current differs from \eqref{cus:1} by the presence of two additional terms which depend on the acceleration.  However, when integrated over the whole three-space as in \eqref{cus:2}, $J^{\rm LW}_\MU$ yields the same total current $e\dot{Z}_\MU$ as $J^{\rm S}_\MU$, because after multiplication by the volume element these acceleration dependent terms do not contribute to the radial integral since $\xi\delta(\xi)=0$.  Therefore, as distributions, the current densities \eqref{cus:1} and \eqref{cus:5} are equivalent. 

  Unfortunately, if one tries to verify that the Li\'enard-Wiechert field charge-current density $J^{\rm LW}_\MU$ is locally conserved, and consistently uses again the differentiation rule \eqref{def:6}, one finds
\begin{align}\label{cus:6}
   \partial^\MU J^{\rm LW}_\MU = \frac{e}{4\pi}
                           2 \kappa \frac{1}{\xi^2}\delta(\xi) \neq 0.
\end{align}
This result is not identically zero as it should be according to \eqref{int:1}.  Also it is distributionally non-zero and well-defined because $\int \rmd\xi\,\xi^2 \partial^\MU J^{\rm LW}_\MU$ is finite.  The current density $J^{\rm LW}$ is therefore not locally conserved.  Moreover, if instead of $J^{\rm LW}_\MU$ one tries to verify local conservation for $J^{\rm S}_\MU$ defined by \eqref{cus:1} one also finds that it is not locally conserved, i.e.,
\begin{align}\label{cus:7}
   \partial^\MU J^{\rm S}_\MU = - \frac{e}{4\pi}
                           2 \kappa \frac{1}{\xi^2}\delta(\xi) \neq 0.
\end{align}
Therefore, even if the additional term in $J^{\rm LW}_\MU$ is discarded to get $J^{\rm S}_\MU$, what is the usual practice in distribution theory, the divergence is still non-zero.  Only when $\kappa=0$, i.e., for non-accelerated motion, are these two current densities locally conserved.  This means that something is wrong in the customary formulation of the electrodynamics of an arbitrarily moving point-charge, or else that something is mathematically inconsistent and needs to be clarified.

\section{The locally conserved current}
\label{loc:0} \setcounter{equation}{0}

Before explaining the reasons of the non local-conservation of the current $J^{\rm LW}_\MU$, let us find out under which conditions the potential \eqref{cus:3} leads to a locally conserved current.  In view of this we remark that any current distribution of the general form
\begin{align}\label{loc:1}
    J_\MU(X)= \frac{e}{4\pi} \Bigl( \frac{\dot{Z}_\MU}{\xi^2} + \frac{S_\MU}{\xi} + T_\MU
             \Bigr)\delta(\xi),
\end{align}
where $S_\MU$ and $T_\MU$ are any smooth four-vector functions, will satisfy global charge conservation because of the distributional identity $\xi\delta(\xi)=0$.  We therefore suppose that the origin of the absence of local current conservation could be due to an incorrect handling of the singularity at $\xi=0$, which should in fact lead to a current having a more complicated form than \eqref{cus:1} or \eqref{cus:5}.

  To proceed step by step and be fully general, we begin by replacing the $1/\xi$ factor in the potential \eqref{cus:3} by a function $\varphi(\xi)$ that is finite and differentiable a sufficient number of times $n$, except possibly at $\xi=0$, i.e.,
\begin{align}\label{loc:2}
    A_\MU \DEF e \dot{Z}_\MU(\tau) \varphi(\xi) \Bigr|_{\tau=\tau_r},
  \quad \text{where} \qquad
   \varphi \in \mathcal{C}^n, \forall \xi \neq 0.
\end{align}
  Moreover, in order to make gauge invariance explicit, we write the equation for the current density $J_\MU$ directly in terms of $A_\MU$, i.e.,
\begin{align}\label{loc:3}
       4\pi J_\MU
     = \partial_\NU \partial^\NU A_\MU 
     - \partial_\MU \psi,
     \qquad\mbox{where}\qquad
     \psi = \partial^\NU A_\NU,
\end{align}
is the invariant scalar which is set equal to zero in the Lorentz gauge. After a long but elementary calculation we find 
\begin{align}
\label{loc:4} 
     \psi = e \kappa \xi \varphi_1(\xi),
     \qquad\mbox{where}\qquad
       \varphi_1 \DEF \varphi' + \frac{1}{\xi}\varphi,\\
\label{loc:5} 
     \partial_\NU \partial^\NU A_\MU =
     e 2 \ddot{Z}_\MU \varphi_1(\xi)
   - e (1-2\kappa\xi) \dot{Z}_\MU (\varphi_1' + \frac{1}{\xi}\varphi_1),\\
\label{loc:6} 
     \partial_\MU \psi = 
     e (\ddot{Z}_\MU + \xi \chi K_\MU) \varphi_1(\xi)
     + e\kappa\xi\Bigl( (1-\kappa\xi)K_\MU + \dot{Z}_\MU \Bigr) \varphi_1'(\xi).
\end{align}
By a tedious calculation we can then explicitly verify that $\partial^\MU J_\MU \equiv 0$ when $Z_\MU(\tau)$ and $\varphi(\xi)$ are any three times differentiable functions, provided $\xi \neq 0$.

   The next step is then to take for $\varphi(\xi)$ a distribution rather than a $\mathcal{C}^n$ function, and to require that the current density $J_\MU$ must still be locally conserved, even if $\xi = 0$.  Specifically, we consider $\varphi=1/\xi$ for which
\begin{align}\label{loc:7} 
  \forall \xi > 0,  \qquad 
  \varphi_1(\xi) = -\frac{1}{\xi^2} + \frac{1}{\xi^2} = 0,
\end{align}
so that that equations \eqref{loc:4} to \eqref{loc:6} are all zero when $\xi \neq 0$.  The current density $J_\MU$ is thus everywhere zero, except at $\xi =0 $ where it is undefined. We therefore interpret $\varphi_1(\xi)$ as a distribution, and use the theorem stating that \emph{a distribution which has its support only in one point, say the origin, is a linear combination of the $\delta$-function and its derivatives up to a certain order} \cite[p\,784]{COURA1962-}.  Thus
\begin{align}\label{loc:8} 
  \forall \xi \geqslant 0,  \qquad \varphi_1(\xi)
      = \frac{1}{\xi} \delta(\xi),
\end{align}
which because of dimensionality comprises a single $\delta$-function, and whose  normalization will turn out to be consistent with \eqref{cus:2}.  It remains to substitute this expression in  \eqref{loc:5} and \eqref{loc:6}, and the locally conserved current-density \eqref{loc:3} is finally found to be 
\begin{align}\label{loc:9}
    J_\MU=  \frac{e}{4\pi}
             \Bigl(   \frac{\dot{Z}_\MU}{\xi^2} 
                    + \frac{\ddot{Z}_\MU + 2\kappa K_\MU}{\xi}
                    - (2\kappa^2 + \chi) K_\MU
              \Bigr)  \delta(\xi)\Bigr|_{\tau=\tau_r}.
\end{align}

   This leads to several observations:
\begin{enumerate}

\item The structure of the current density $J_\MU$ is much more complicated than that of the customary current \eqref{cus:1}.  $J_\MU$ depends directly on the three invariants $\xi, \kappa$, and $\chi$, as well as on the two four-vectors $\dot{Z}_\MU$ and $\ddot{Z}_\MU$; indirectly on the biacceleration $\dddot{Z}_\MU$ through the invariant $\chi$; and, finally, on the angular variables through the null four-vector $K_\MU(\theta,\phi)$.

\item The dependence on the third derivative of $Z_\MU$ is consistent with the Lorentz-Dirac equation and with the Schott expression of the self-force, in which $\dddot{Z}_\MU$ also appears, because the self-interaction force involves a product of $J_\MU$ with the self-field.

\item Equation \eqref{loc:9} has the most general distributional form of \eqref{loc:1}, in accord with the theorem cited above \eqref{loc:8}.

\item The equation $\varphi_1(\xi) = \varphi' + \varphi/\xi = 0, \forall \xi \neq 0,$ has only one solution: $1/\xi$. This singles out the corresponding potential as being the only one such that the current density of a point-charge is conserved and thus given by \eqref{loc:9}.

\end{enumerate}

\section{Straightforward derivation}
\label{str:0} \setcounter{equation}{0}

While the derivation made in the previous section is rigorous, it is indirect in the sense that it gives no explanation for the origin of the $\delta$-functions, which like any distribution must by Schwarz's structure theorem come from the partial differentiation of some continuous function.  In fact, this generating function is easily found because in three-dimensional notation the retarded distance \eqref{def:2} reads
\begin{align}\label{str:1}
              \xi = |\vec{x} - \vec{z}\,| \gamma(1-\vec{\rho}\cdot\vec{\beta}),
\end{align}
where $\vec{\rho}$ is the unit vector in the direction of $\vec{x} - \vec{z}$. The retarded distance is therefore proportional to an absolute value, and for that reason has a discontinuous derivative when $\vec{x} \rightarrow \vec{z}$, i.e., at $\xi=0$ where a second partial differentiation leads to a $\delta$-function.

   Consequently, as a direct generalization of the case of a point-charge at rest, which is discussed in details in \cite{GSPON2004D,GSPON2006B,GSPON2008B}, the potential of an arbitrarily moving accelerated point-charge can be written
\begin{align}\label{str:2}
    A_\MU = e \frac{\dot{Z}_\MU}{\xi}\ \Upsilon(\xi) \Bigr|_{\tau=\tau_r},
\end{align}
where $\Upsilon(\xi)$ is the generalized function defined as\footnote{Intuitively, $\Upsilon(r)$ can be seen as equivalent to the sign function ${\rm sgn}(r)$ for $r \geq 0$.}
\begin{align} \label{str:3}
   \Upsilon(r) \DEF 
         \begin{cases}
         \text{undefined}   &   r < 0,\\
                      0     &   r = 0,\\
                     +1     &   r > 0,
         \end{cases}
   \qquad\text{and}\qquad
     \frac{d}{dr}\Upsilon(r) = \delta(r),
\end{align}
which explicitly specifies how to consistently differentiate at $\xi=0$.  

   When the definition \eqref{def:5} and the causal differentiation rule \eqref{def:6} are now used to calculate the field strength starting from the potential \eqref{str:2}, the corresponding current-density \eqref{int:2} is directly found to be the conserved one, i.e., \eqref{loc:9}.  However, instead of the customary Li\'enard-Wiechert field \eqref{cus:4}, the field strength is now
\begin{align}
\label{str:4}
   F_{\mu\nu} &=  e \Bigl( 
            \frac{K_\MU\ddot{Z}_\NU}{\xi}
            + (1-\kappa\xi)\frac{K_\MU \dot{Z}_\NU}{\xi^2}
            - \{ \mu \leftrightarrow \nu \}
                   \Bigr)\Upsilon(\xi) \Bigr|_{\tau=\tau_r}\\
\label{str:5}
               &- e \Bigl( 
              (1-\kappa\xi)\frac{K_\MU \dot{Z}_\NU}{\xi}
            - \{ \mu \leftrightarrow \nu \}
                   \Bigr)\delta(\xi) \Bigr|_{\tau=\tau_r},
\end{align}
which apart from the presence of the $\Upsilon$-function multiplying $F_{\mu\nu}^{LW}$ on the first line, has an additional $\delta$-like contribution on the second line.  This field is the proper relativistic generalization of the Coulomb-Tangherlini field of a point-charge \cite{GSPON2004D,GSPON2006B,GSPON2008B}.  Since both the $\Upsilon$-factor and the $\delta$-like contribution are necessary to obtain the current density satisfying the local conservation identity \eqref{int:1}, it becomes clear why the customary $F_{\mu\nu}^{\rm LW}$ cannot lead to such a current.  In fact, it is by calculating the current density immediately from the potential as in \eqref{loc:3} --- that is by ignoring that the field could be different from the customary one --- that after many unsuccessful attempts the author discovered the conserved current-density in July 2003.

  The implication of the necessity of the $\delta$-like contribution \eqref{str:5} to get an identically conserved charge-current density is that the field (\ref{str:4}--\ref{str:5}) cannot be interpreted as a distribution in which \eqref{str:5} can be ignored if the field is twice differentiated to calculate the divergence \eqref{int:1}.  The reason is that every successive differentiation introduces new terms which must be retained in order to get a result that is identically zero.\footnote{This can be illustrated by the following example: the function $f(\xi)=\xi^{-1}\delta(\xi)$ yields zero when evaluated on any test function $T(\xi)$ in $\mathbb{R}^3$.  On the other hand, $f'(\xi)=-2\xi^{-2}\delta(\xi)$ yields $-2T(0)\neq 0$ so that differentiation and evaluation on test functions do not commute in general.}  The field (\ref{str:4}--\ref{str:5}) and the current density \eqref{loc:9} must therefore be interpreted as nonlinear generalized function, such as Colombeau functions \cite{GSPON2006B}, in which all terms can be or can become significative in further calculations, even though none of the nonlinear properties of such functions are used in the present letter.

\section{Discussion}
\label{dis:0} \setcounter{equation}{0}

In this letter we have derived the proper formulation of the potential of an arbitrarily moving point-charge \eqref{str:2}, which leads to the conserved current-density \eqref{loc:9}, which, most probably, has never been published.  This means that there is an apparent contradiction between this new result and the fact that the customary Li\'enard-Wiechert formulation is an agreement with so many applications of classical electrodynamics.

There is however no contradiction, since, on the contrary, the results of this letter are in full agreement with the fundamental principles of electrodynamics and mechanics.   For instance, if the conserved current \eqref{loc:9} is introduced in an action integral as a scalar product $J_\MU A^\MU_{\rm ext}$ with the potential of an external field $A^\MU_{\rm ext} \in \mathcal{C}^\infty(\mathbb{R}^4)$, the differences between that current and the simple current $J_\MU^{\rm S}$ defined by \eqref{cus:1} have in general no influence since they disappear upon integration over the whole space.  The same is true for the calculation (which also involves an integration) of the Li\'enard-Wiechert potential, i.e., \eqref{cus:2} or \eqref{str:2}, by means of a Green's function. 

Thus, the principles of physics imply that the position $Z_\MU$ and velocity $\dot{Z}_\MU$ of a point-charge are sufficient to determine the potential of its field, while the precise formulation of that potential given by \eqref{str:2} is necessary to determine the complete field and conserved current-density, which include terms that are function of $\ddot{Z}_\MU$ and $\dddot{Z}_\MU$.  In other words, while $J_\MU^{\rm S}$ is sufficient as a source to determine uniquely the potential of an arbitrarily moving point-charge, the conserved current $J_\MU$ deriving from this potential can be very different from $J_\MU^{\rm S}$.  Indeed, without the factor $e$, $J_\MU^{\rm S}$ is merely a velocity distribution associated to the world-line of the point-charge.

In conclusion, the formulation presented in this letter will make little difference in most engineering-type applications of classical electrodynamics.  Indeed, as can be seen by studying a number of examples, the instances in which the full details of the current density \eqref{loc:9} are strictly necessary, and the additional contribution \eqref{str:5} to the field is essential, include fundamental problems like calculating the interaction of a point-charge with itself \cite{GSPON2008B}, and similar problems in which classical electrodynamics is apparently not consistent.  For example, the self-force on an arbitrarily moving point-charge is given by $F_{\mu\nu}  J^\NU$ where $F_{\mu\nu}$ is given by (\ref{str:4}--\ref{str:5}) and $J^\NU$ is the full conserved current density \eqref{loc:9}, i.e., such that $ \partial^\nu F_{\mu\nu} =  - 4\pi J_\MU$, not the customary charge-current density \eqref{cus:1}:  a frequent confusion which is a cause of known difficulties.  The resolution of such internal contradictions is the subject of several forthcoming publications \cite{GSPON2007A,GSPON2007B}.

\section{References}
\label{biblio}


\begin{enumerate}

\bibitem{GSPON2006D} A. Gsponer, \emph{Derivation of the potential, field, and locally-conserved charge-current density of an arbitrarily moving point charge} (2006) 19~pp.  e-print arXiv:physics/0612232.

\bibitem{SOMME1951-} A. Sommerfeld, Electrodynamics (Academic Press, 1948, 1960) 371~pp.

\bibitem{LANDA1975-} L.D. Landau and E.M. Lifshitz, The Classical Theory of Fields (Pergamon Press, 1951, 1975) 402~pp.

\bibitem{JACKS1975-} J.D. Jackson, Classical Electrodynamics (J. Wiley \& Sons, New York, second edition, 1962, 1975) 848~pp.

\bibitem{BARUT1964-} A.O. Barut, Electrodynamics and the Classical Theory of Fields and Particles (Dover, 1964, 1980) 235~pp.

\bibitem{INGAR1985-} R.S. Ingarden and A. Jamiolkowski, Classical Electrodynamics (Elsevier, 1985) 349~pp.

\bibitem{COURA1962-} R. Courant and D. Hilbert, Methods of Mathematical Physics {\bf 2} (Interscience Publ., New York, 1962) 830~pp.


\bibitem{GSPON2004D} A. Gsponer, \emph{Distributions in spherical coordinates with applications to classical electrodynamics}, Eur. J. Phys. {\bf 28} (2007) 267--275; Corrigendum Eur. J. Phys. {\bf 28} (2007) 1241. e-print arXiv:physics/0405133.

\bibitem{GSPON2006B} A. Gsponer, \emph{A concise introduction to Colombeau generalized functions and their applications to classical electrodynamics}, Eur. J. Phys. {\bf 30} (2009) 109--126. e-print arXiv:math-ph/0611069.

\bibitem{GSPON2008B} A. Gsponer, \emph{The classical point-electron in Colombeau's theory of generalized functions}, J. Math. Phys. {\bf 49} (2008) 102901 \emph{(22 pages)}. e-print arXiv:0806.4682.


\bibitem{GSPON2007A} A. Gsponer, \emph{The self-interaction force on an arbitrarily moving point-charge and its energy-momentum radiation rate: A mathematically rigorous derivation of the Lorentz-Dirac equation of motion} (2008) 17~pp. e-print arXiv:0812.3493.

\bibitem{GSPON2007B} A. Gsponer, \emph{Derivation of the self-interaction force on an arbitrarily moving point-charge and of its related energy-momentum radiation rate: The Lorentz-Dirac equation of motion in a Colombeau algebra} (2008) 35\,pp. e-print arXiv:0812.4812.


\end{enumerate}
\end{document}